\documentclass{INTERSPEECH2023}

\interspeechcameraready

\title{Vocoder drift in x-vector--based speaker anonymization}
\name{Michele Panariello, Massimiliano Todisco, Nicholas Evans}
\address{EURECOM, Sophia Antipolis, France}
\email{firstname.lastname@eurecom.fr}

\usepackage{comment}
\usepackage[textsize=scriptsize]{todonotes} 
\usepackage{enumitem}
\usepackage[framemethod=tikz]{mdframed} 
\usepackage{subfig} 
\usepackage{multirow}
\usepackage{makecell}
\usepackage{tabularray}
\SetTblrInner{rowsep=0pt} 
\usepackage{svg}
\usepackage{adjustbox} 
\usepackage{floatrow}
\newfloatcommand{capbtabbox}{table}[\captop][\Xhsize] 
\usepackage{balance}

\newcommand{\signal}{\mathbf{s}}
\newcommand{\xv}{\mathbf{x}}
\newcommand{\xvdim}{m}

\newcommand{\xvo}{\mathbf{x}_o}
\newcommand{\odom}{\hat{O}}

\newcommand{\xvpre}{\mathbf{x}_p}
\newcommand{\predom}{\hat{P}}

\newcommand{\xvpost}{\mathbf{x}_a}
\newcommand{\postdom}{\hat{A}}

\newcommand{\anon}[1]{a\left(#1\right)}
\newcommand{\voc}[1]{v\left(#1\right)}
\newcommand{\anonvoc}[1]{v \circ a : #1}

\newcommand{\target}{\emph{target}}
\newcommand{\drift}{\emph{drift}}

\newcommand{\secref}[1]{Section~\ref{#1}}
\newcommand{\figref}[1]{Figure~\ref{#1}}
\newcommand{\tabref}[1]{Table~\ref{#1}}

\newcommand{\attention}[1]{#1}

\begin{document}

\maketitle
 
\begin{abstract}
State-of-the-art approaches to speaker anonymization typically employ some form of perturbation function to conceal speaker information contained within an x-vector embedding,
then resynthesize utterances in the voice of a new pseudo-speaker using a vocoder.
Strategies to improve the x-vector anonymization function have attracted considerable research effort, whereas vocoder impacts are generally neglected. In this paper, we show that the impact of the vocoder is substantial and sometimes dominant. The vocoder drift, namely the difference between the x-vector vocoder input and that which can be extracted subsequently from the output, is learnable and can hence be reversed by an attacker;
anonymization can be undone and the level of privacy protection provided by such approaches might be weaker than previously thought. The findings call into question the focus upon x-vector anonymization, prompting the need for greater attention to vocoder impacts and stronger attack models alike.
\end{abstract}

\section{Introduction}
\label{sec:introduction}
The task of \emph{speaker anonymization} broadly refers to the processing of speech recordings to conceal the speaker identity while preserving the linguistic and paralinguistic content. 
The topic has attracted increasing research interest in recent years, in particular through the VoicePrivacy Challenge~\cite{vp20_findings, vp22_eval}, which was founded in 2020 to define the problem, provide strong baselines, foster progress and identify research priorities.
No matter what the application, anonymization should protect an appropriate
trade-off between \emph{privacy} and \emph{utility}.
Privacy can be estimated using automatic speaker verification (ASV) and an equal error rate (EER) metric to gauge the ability of an attacker to infer the true speaker identity. 
Utility is estimated 
using automatic speech recognition (ASR) and a word error rate (WER) metric which reflects the degree to which linguistic and paralinguistic content is preserved.

Most anonymization solutions are based upon original work~\cite{Fang2019} and
upon the extraction and processing of three different representations~\cite{xv_anon}:
\begin{itemize}[noitemsep, nolistsep]
    \item a set of linguistic features produced by an ASR model;
    \item a representation of intonation and prosody, usually in the form of a fundamental frequency (F0) curve;
    \item an \emph{x-vector}, namely
    a neural embedding which encodes the speaker identity~\cite{xvectors_original_paper}.
\end{itemize}
To conceal the speaker identity, the x-vector is typically perturbed by means of an anonymization function, thereby obtaining a new \emph{pseudo-speaker} embedding. 
The three components are then fed to a waveform synthesis model (a \emph{vocoder}) to produce an 
utterance in the voice of
the pseudo-speaker.

The anonymization function used by two of the three VoicePrivacy baselines utilizes a \emph{pool} of external x-vectors.
The pseudo-speaker x-vector is derived from a subset of the furthest vectors in the pool from the input x-vector.
Most VoicePrivacy participants focused predominantly upon improving the anonymization function
to enhance privacy~\cite{participant1, participant2, participant3}.
This focus can imply an assumption that no other processing stages contribute substantially to anonymzation.
We have found this not to be the case.

We report in this paper our work to observe and compare the relative impacts of a conventional x-vector anonmyization function and a vocoder, two components of a state-of-the-art anonymization system~\cite{miao22_odyssey}.
We show that both components contribute to anonymization and that the contribution of the vocoder, which we refer to as the \emph{vocoder drift}, is in some cases even greater than that of the anonymization function.
We demonstrate that this phenomenon is also common to other popular vocoders. 
Collectively, they fail to provide the level of fine-grained control over the input/output x-vector space that would otherwise justify the focus within the community upon the anonymization function.
Finally, we show that the vocoder drift can be learned by an attacker, knowledge which can be exploited in order to reverse the anonymization.
Our findings corroborate other evidence~\cite{pierre_attack} that
the protection provided by such approaches to anonymization might be overestimated.

\section{Relation to prior work}
\label{sec:prior_work}
\begin{figure}[!t]
    \centering
    \includegraphics[width=\columnwidth]{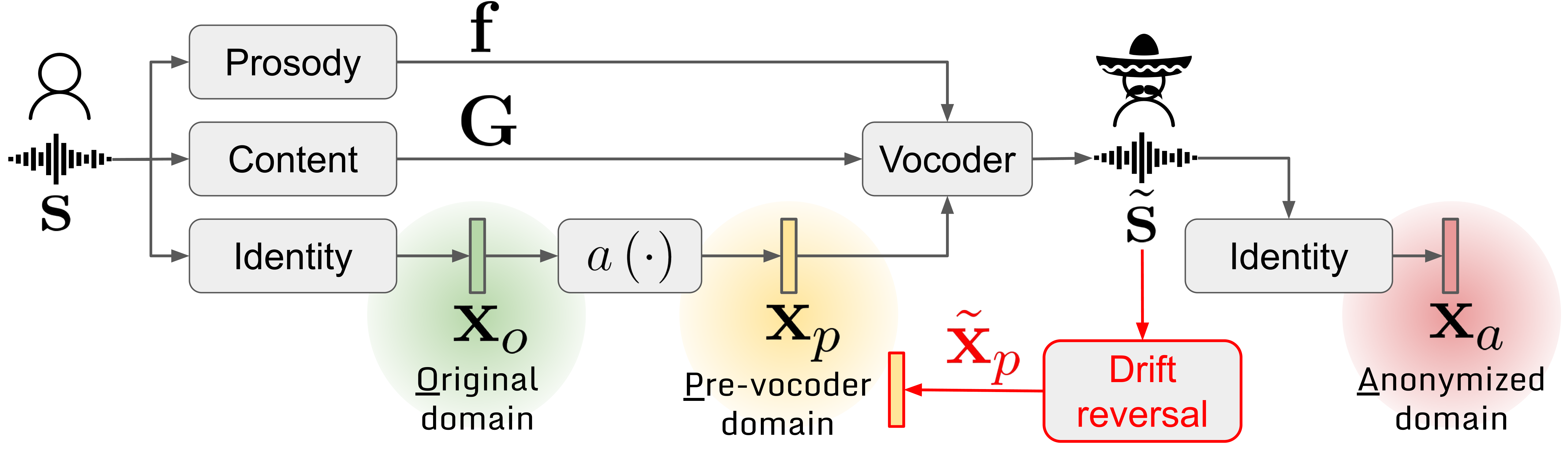}
    \caption{Overview of a conventional speaker anonymization system and the different x-vector domains. The block in red represents the vocoder drift reversal attack reported in \secref{sec:attack_on_vocoder}.}
    \label{fig:anon_system}
\end{figure}

In this section we describe the typical, high-level structure of an x-vector--based speaker anonymization system (see \figref{fig:anon_system}), along with relevant prior work.
We then
introduce our own setup which we used for all experiments reported in \secref{sec:drift} and \secref{sec:attack_on_vocoder}.

\subsection{X-vector--based speaker anonymization}
Let $\signal \in \mathbb{R}^L$ be an input speech utterance of $L$ samples.
The input is first frame-blocked into a sequence of $N$ frames and then decomposed
into three separate representations comprising: an F0 curve $\mathbf{f}\in\mathbb{R}^{N}$ which is intended to encode intonation and prosody; a set of $c$-dimensional linguistic features $\mathbf{G}\in\mathbb{R}^{c \times N}$ which encode the spoken content (the text); an x-vector $\xvo \in \mathbb{R}^\xvdim$ which encodes the speaker identity, where subscript $o$ denotes extraction from an \emph{original} input utterance.

A vocoder model $V(\mathbf{f}, \mathbf{G}, \xvo)$ is trained to reconstruct input waveforms from the decomposition.
Anonymization is achieved by replacing $\xvo$ with a substitute so as to conceal the speaker identity, but by using the other components unchanged in order to preserve remaining speech attributes.
The substitution is performed using an 
anonymization function
$\anon{\xvo} = \xvpre \in \mathbb{R}^\xvdim$ to perturb the original x-vector.
An anonymized utterance $\Tilde{\signal}$ in the voice of a fictitious, pseudo-speaker determined by the anonymized x-vector $\xvpre$ is then synthesized according to $\Tilde{\signal} = V(\mathbf{f}, \mathbf{G}, \xvpre)$.
The anonymized utterance should maintain the same linguistic and paralinguistic content as the original input signal.
As discussed later, an additional x-vector $\xvpost$ can be extracted from $\Tilde{\signal}$ in order to measure privacy.

By convention, $\anon{\cdot}$
acts to create a new pseudo-speaker using speaker embeddings drawn from an external pool of x-vectors~\cite{vp20_findings,vp22_eval,Fang2019,xv_anon,miao22_odyssey,pierre_attack}.
Given an input $\xvo$, the $K$ vectors within the pool that are furthest from $\xvo$
according to some distance metric
are selected and then, from among them, $K^*$ vectors are chosen randomly and averaged to obtain $\xvpre$.
The design of this function has received considerable attention, with numerous works having investigated 
how its configuration, the choice of distance metric~\cite{ablation} and the strategy by which x-vectors are selected from the pool~\cite{ablation, champion_study} influence performance.
The participants of the two VoicePrivacy Challenges held in 2020 and 2022 proposed different enhancements to $\anon{\cdot}$.
They include the generation of pseudo-speaker embeddings using a generative adversarial network~\cite{participant1, gan}
and adversarial noise~\cite{participant2}, among others~\cite{participant3,participant4}.
None of the participants reported the influence of the vocoder.
In this paper, we show that it too contributes to anonymization and that it can be
responsible for a great deal
of the privacy protection.

\subsection{Our approach}
\label{sec:experimental_setup}
Our approach is based on the pipeline described in~\cite{miao22_odyssey}.\footnote{Code available at \url{github.com/eurecom-asp/vocoder-drift}.}
The F0 curve is estimated using YAAPT~\cite{yaaaaaaapt}. The linguistic feature extractor is a HuBERT-based soft content encoder~\cite{hubert} and x-vectors are extracted using  \mbox{ECAPA-TDNN}~\cite{ecapa}.
We experimented with three vocoders:
the HiFi-GAN~\cite{hifi_gan}, originally used in~\cite{miao22_odyssey};
the neural source filter (NSF) model~\cite{nsf} as used by baseline B1a of the VoicePrivacy Challenge held in 2022;
a variation of the HiFi-GAN which uses a NSF model as generator, as used by baseline B1b of the same VoicePrivacy Challenge edition~\cite{vp22_eval}.
We use the conventional pool-based anonymization function $\anon{\cdot}$ described above with $K=200$, $K^*=100$, and a cosine distance metric.

Like most related work, we use the VoicePrivacy database and standard protocols~\cite{vp22_eval}.
The \textit{LibriTTS-train-clean-100} dataset is used for vocoder training.
The \textit{LibriSpeech-test-clean} and \textit{VCTK} datasets (decomposed into male and female subsets) are used for evaluation.
The external pool of x-vectors is derived using the \mbox{\textit{LibriTTS-train-other-500}}~\cite{libriTTS} dataset.
Privacy is evaluated
using ASV experiments comprising a set of enrollment utterances that an attacker attempts to match to a set of protected (anonymized) trial utterances.
ASV is performed by scoring x-vectors with the cosine distance and without any additional backend processing~\cite{miao22_odyssey, ecapa}.

\section{Vocoder drift}
\label{sec:drift}
In this section, we 
introduce the notion of \emph{vocoder} \drift{} and report an investigation of its impact 
upon x-vector pertubation and privacy.  
\label{sec:drift_general}

\subsection{Definition}

Figure~1 shows the three x-vectors used in this work. The first $\xvo$ is extracted from the original utterance $\signal$ (left in Figure~1).  
A second x-vector $\xvpost$ can be extracted from 
the anonymized utterance $\Tilde{\signal}$ (right).
The third x-vector $\xvpre$ is the output of the anonymization function (middle).
We denote the separate domains of $\xvo$, $\xvpost$ and $\xvpre$ as $\odom$ (original), $\postdom$ (anonymized) and $\predom$ (pre-vocoder), respectively.

As described in \secref{sec:prior_work}, the majority of research
has focused on improving the anonymization function $\anon{\cdot}$, the general hypothesis being that this component is primarily responsible for ensuring privacy. 
Intuitively, privacy is improved by 
increasing the difference between $\xvo$ and $\xvpre$, e.g.\ according to the cosine distance.
With the focus being upon the anonymization function, there is an inherent, perhaps unrealistic assumption that the vocoder preserves this distance such that the difference, which we term as the \emph{drift}, between the x-vectors at the input ($\xvpre$) and that which can be extracted from the output ($\xvpost$) is only modest. 
In this work, we seek to test this assumption.

We model the relationship between the $\predom$ and $\postdom$ domains with a function $\voc{\xvpre} = \xvpost$. 
It allows us to define the trajectory of an x-vector through the whole anonymization system as $\anonvoc{\xvo \mapsto \xvpost}$, where $\circ$ denotes function composition.

\subsection{X-vector perturbation}
\label{sec:drift_def}
In seeking to quantify the impact of $\voc{\cdot}$ on the \mbox{x-vector} trajectory,
we define two metrics. Let $d$ be some distance measure over $\mathbb{R}^\xvdim$.  We then define:
\begin{itemize}[noitemsep, nolistsep]
    \item $d(\xvo,\xvpre)$ as the \emph{target} distance, a measure of how far $\xvo$ is perturbed away from its original position according to $\anon{\cdot}$;
    \item $d(\xvpre, \xvpost)$ as the vocoder \emph{drift}, a measure of the shift between the input x-vector $\xvpre$ and that which can be extracted from the vocoder output $\xvpost$, introduced by means of $\voc{\cdot}$.
\end{itemize}
Intuitively, it is desirable that \drift{} $\ll$ \target{}, which means the anonymization system provides fine-grained control over the final position of $\xvpost$: it is close to the targeted pseudo-speaker embedding $\xvpre$.
If this is not the case, then the x-vector trajectory is determined in considerable part by $\voc{\cdot}$; the x-vector extracted from the output $\xvpost$ is far from the target and the system does not provide fine-grained control over the x-vector space in $\postdom$.

\begin{table}[!t]
\caption{Average target distance and drift for each vocoder and each test set of LibriSpeech and VCTK, separated by speaker sex. All cosine distances have a standard deviation between $0.05$ and $0.10$.}
\label{tab:drift}
\centering
\resizebox{\columnwidth}{!}{%
\footnotesize
\begin{tblr}{lcccc}
\hline[1pt]
\SetCell[r=2]{} & \SetCell[r=2]{} \textbf{target} & \SetCell[c=3]{} \textbf{drift} \\
\cline{3-5} & & HiFi-GAN & \rule{0pt}{8pt} NSF & HiFi-NSF \\ \hline 
LibriSpeech (F) & $1.3$ & $0.62$ & $0.91$ & $0.97$ \\
LibriSpeech (M) & $1.2$ & $0.56$ & $0.80$ & $0.94$ \\
VCTK (F) & $1.3$ & $0.67$ & $0.92$ & $0.94$ \\
VCTK(M) & $1.3$ & $0.59$ & $0.90$ & $0.90$ \\ \hline[1pt]
\end{tblr}}
\end{table}

We compute the average \drift{} and \target{} for each database subset and each vocoder: results are shown in \tabref{tab:drift}.
The \target{} is in the order of 1.3 for all four subsets.
The value of these distances lies in their \emph{comparison} to estimates of the \drift{} shown in the last three columns. 
For the HiFi-GAN vocoder, the drift is almost half the target distance. 
Lying between 0.8 and 0.97, the drift for the NSF and HiFi\nobreakdash-NSF vocoders is substantially greater still, with drift distances almost as large as target distances. 
These results show that the control over the x-vector domain $\postdom$ is potentially low and
suggest that the x-vector anonymization and vocoder functions have an almost-comparable contribution to x-vector perturbation.
It is still necessary, however, to explore their resulting impact upon \emph{privacy}.

\subsection{Impacts upon privacy}
\label{sec:drift_priv}
We follow the VoicePrivacy-defined approach to measure privacy impacts.
We report a set of ASV experiments using different combinations of x-vectors.
In all cases,
privacy is measured using estimates of the EER.
Enrollment and trial utterances are as defined by the VoicePrivacy protocol (see \secref{sec:experimental_setup}).  
There are several enrollment utterances per speaker. 
Individual x-vectors are extracted from each, averaged,
and compared to a number of trial utterances.
For each utterance, we extract the set of~$\xvo$, $\xvpre$ and $\xvpost$ x-vectors.
Each set of experiments is conducted three times, with each iteration using one of the three different x-vectors.
Results using the set containing $\xvo$
x-vectors \attention{($\odom$ domain)} provide a baseline. 
Those derived from the set of $\xvpre$ x-vectors \attention{($\predom$ domain)} provide an indication of the contribution to privacy of the anonymization function~$\anon{\cdot}$.
Results using final set containing $\xvpost$ x-vectors \attention{($\postdom$ domain)} provide an indication of the contribution of the vocoder function~$\voc{\cdot}$.
Once again, we report results for the same experiment using all three vocoders.

\begin{table}[!t]
  \caption{Privacy protection of the x-vector domains at different stages of the anonymization pipeline (EER, \%) on test sets of LibriSpeech and VCTK, separated by speaker sex.}
  \label{tab:privacy}
\centering
\resizebox{\columnwidth}{!}{%
\begin{tblr}{lclccc}
\hline[1pt]
\SetCell[r=2]{} & \SetCell[r=2]{} $\odom$ dom. & \SetCell[r=2]{} $\predom$ dom. & \SetCell[c=3]{} $\postdom$ dom. \\ \cline{4-6} 
 & & & \rule{0pt}{8pt} HiFi-GAN & NSF & HiFi-NSF \\ \hline
LibriSpeech (F) & $0.54$ & $2.51$ & $15.0$ & $17.9$ & $16.2$ \\
LibriSpeech (M) & $0.88$ & $2.99$ & $14.5$ & $20.3$ & $19.0$ \\
VCTK (F) & $1.13$ & $5.59$ & $25.3$ & $31.0$ & $28.1$ \\
VCTK(M) & $0.17$ & $3.04$ & $18.5$ & $16.7$ & $19.1$ \\ \hline[1pt]
\end{tblr}}
\end{table}

\begin{figure*}
\begin{floatrow}
\floatbox[]{table}[\FBwidth][][c]
{
    \caption{Performance of the proposed drift-reversal attack compared to a lazy-informed attack and a supervised semi-informed attack (EER, \%) on the LibriSpeech and VCTK test sets.}\label{tab:attack}
}
{
    \resizebox{0.55\textwidth}{!}{ 
        \begin{tblr}{Q[c,m]Q[c,m]Q[c,m]Q[c,m]Q[c,m]Q[c,m]}
    \hline[1pt]
    \textbf{Vocoder} & \textbf{Dataset} & \textbf{Unprotected} & {\textbf{Lazy}\\\textbf{informed}} & {\textbf{Semi}\\\textbf{informed}} & {\textbf{Drift}\\\textbf{reversal}} \\
    \hline
    \SetCell[r=4]{} HiFi-GAN & LibriSpeech (F) & $0.54$ & $11.3$ & $3.21$ & $\mathbf{3.10}$ \\
    & LibriSpeech (M) & $0.88$ & $10.9$ & $\mathbf{1.78}$ & $4.45$ \\
    & VCTK (F) & $1.13$ & $19.2$ & $13.3$ & $\mathbf{7.53}$ \\
    & VCTK (M) & $0.17$ & $11.0$ & $8.14$ & $\mathbf{3.79}$ \\
    \hline
    \SetCell[r=4]{} NSF & LibriSpeech (F) & $0.54$ & $15.3$ & $\mathbf{1.92}$ & $5.05$ \\
    & LibriSpeech (M) & $0.88$ & $13.8$ & $\mathbf{1.94}$  & 6.04 \\
    & VCTK (F) & $1.13$ & $24.7$ & $\mathbf{15.7}$ & $16.5$ \\
    & VCTK (M) & $0.17$ & $\mathbf{9.81}$ & $12.3$ & $10.7$ \\
    \hline
    \SetCell[r=4]{} HiFi-NSF & LibriSpeech (F) & $0.54$ & $12.6$ & $\mathbf{4.01}$ & $4.23$ \\
    & LibriSpeech (M) & $0.88$ & $14.7$ & $\mathbf{2.23}$  & $4.90$ \\
    & VCTK (F) & $1.13$ & $22.8$ & $18.4$ & $\mathbf{14.1}$ \\
    & VCTK (M) & $0.17$ & $13.5$ & $11.7$ & $\mathbf{11.1}$ \\
    \hline[1pt]
\end{tblr}
    }
}

\floatbox[]{figure}[\Xhsize][][c]
{
    \caption{t-SNE visualization of the x-vector trajectory for LibriSpeech trial utterances (M) across the three x-vector domains (left). Focus on the trajectory of a single speaker (right). Best viewed in color.}\label{fig:drift}
}
{
  \includegraphics[width=\linewidth]{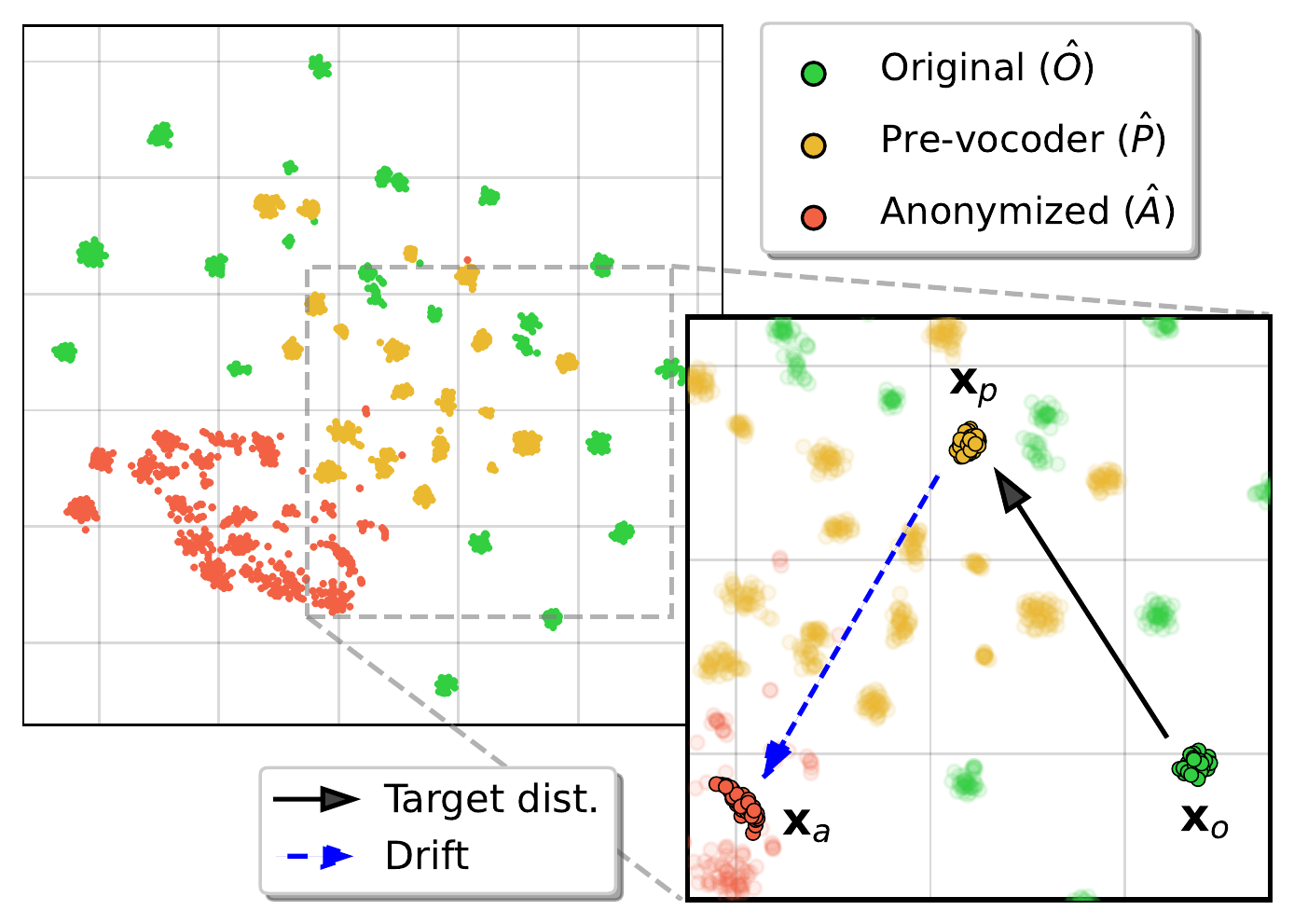}
}
\end{floatrow}
\end{figure*}

Results are shown in \tabref{tab:privacy}, for the same database subsets as in \tabref{tab:drift}. 
Baseline results for the $\odom$ domain show EERs of approximately 1\%.
In the $\predom$ domain, increases in the EER to between 2.5\% and 5.6\% indicate that the anonymization function delivers only a low level of privacy. 
In the $\postdom$ domain, however, EERs are substantially higher for all three vocoders\attention{, if still far from providing perfect privacy (EERs of ~50\%)}. 
The comparison of results for $\predom$ and $\postdom$ domains show that 
the vocoder plays a dominant role; 
most of the anonymization can be attributed to vocoder drift.
We explored this phenomenon with t-SNE visualizations~\cite{tsne} of pooled x-vectors. 
Results are illustrated in \figref{fig:drift}, which depicts a distribution of x-vectors for the male partition of the LibriSpeech dataset. 
In the $\predom$ domain, speaker clusters are still clearly distinguishable, while the bulk of the anonymization can be attributed to vocoder drift.

One could claim that these findings are neither surprising, nor cause for concern.  
There is no guarantee that the
vocoder function $\voc{\cdot}$ 
is invertible in any way which would allow the recovery of x\nobreakdash-vector inputs $\xvpre$ in the $\predom$ domain. 
Since the attacker does not have access to the $\predom$ domain, but only to the $\postdom$ domain, whether anonymization is attributed to the anonymization function or the vocoder function is of little consequence.
In the next section, we disprove these arguments and show that 
an attacker \emph{can} learn this function or, more specifically, how to undo it.
Armed with the inverse function $v^{-1}\left({\cdot}\right)$, an attacker can estimate an x-vector in the $\predom$ domain that corresponds to an x-vector in the $\postdom$ domain and hence 
\emph{reverse} the anonymization.

\section{Drift-reversal attacks}
\label{sec:attack_on_vocoder}

In this section, we introduce \emph{drift reversal}, a novel attack against anonymization systems.

\subsection{Attacks on anonymization systems}
Since speaker anonymization is 
a relatively new research topic, it is hardly surprising that little attention has been dedicated to attacks against it.
Even so, the VoicePrivacy Challenge has explored the robustness of anonymization systems under a so-called \emph{semi-informed} attack model~\cite{vp22_eval}.
Under this scenario, an attacker is aware of anonymization having been performed, and seeks to overcome it (break the anonymization) by using a similar system to generate anonymized data with which to train an ASV system.
Evaluations using ASV systems trained using in-domain (similarly anonymized) data show the potential for attacks to circumvent anonymization.
A more explicit approach is reported in~\cite{pierre_attack} and
can be used by an attacker to invert
a complete anonymization system by means of a rotation matrix and to estimate speaker embeddings $\xvo$ in the unprotected domain $\odom$ from protected x-vectors $\xvpost$ in $\postdom$. 
Our approach is different since we
aim to explore the anonymization robustness when we revert only the \emph{vocoder} \drift{} to recover an estimate of $\xvpre$ in $\predom$.

\subsection{Definition and implementation}
In the case that the bulk of the anonymization performance can be attributed to the vocoder function $\voc{\cdot}$ instead of the anonymization function $\anon{\cdot}$, 
a drift reversal attack can be mounted to undo most of the protection
Let $\signal^{(e)}$ be an original (i.e.\ unprotected) enrollment utterance. 
An attacker can derive a representation of this signal in the $\predom$ domain by extracting an x-vector $\xvo^{(e)}$ and then by computing $a(\xvo^{(e)}) = \xvpre^{(e)}$.
Now let $\Tilde{\signal}^{(t)}$ be an anonymized trial utterance with corresponding x-vector $\xvpost^{(t)}$ in the $\postdom$ domain. 
The attacker can estimate a representation in the $\predom$ domain $\xvpre^{(t)}$ by 
reversing the vocoder drift, i.e.
by computing $v^{-1}(\xvpost^{(t)})$.

While the inverse function is not analytically tractable, the attacker can attempt to \emph{learn} a function $g_\theta(\cdot) \approx v^{-1}(\cdot)$ using a database of training pairs $\xv_{p_i}$ and anonymized utterances $\Tilde{\signal}_i$.
Function $g_\theta$ can be learned using a neural network to map an anonymized utterance $\Tilde{\signal}$ to an approximation of the corresponding x-vector $\xvpre$ in $\predom$. 
This can be achieved by optimizing the objective function
\begin{equation}
    \min_\theta d\left(\xvpre, g_\theta\left(\Tilde{\signal}\right)\right)
\end{equation}
where $d$ is the cosine distance.
Training pairs $\left\{\left(\xv_{p_i}, \Tilde{\signal}_i\right)\right\}_i$ can be obtained by applying anonymization to \emph{any} appropriate (even unlabeled) speech dataset.

Because function $g_\theta$ is effectively an x-vector extraction operation, we fine-tune a pretrained ECAPA-TDNN model
to learn it.
In line with the VoicePrivacy protocol, the model is trained using the \textit{LibriSpeech-train-clean-360} dataset, although approximately 3\% of the data is set aside for validation purposes. 
Still in line with the VoicePrivacy protocol, anonymization is performed at the \emph{speaker level}\footnote{${\xvpre}_i$ is estimated once
for each speaker $i$ and the same ${\xvpre}_i$ is used for each
utterance corresponding to the same speaker -- see~\cite{vp22_eval} for details.} in deriving $\xvpost$ for each enrollment and trial utterance, instead of at the utterance level.
The network is fine-tuned for 3 epochs using Adam optimizer~\cite{adam} with a learning rate of $5 \cdot 10^{-5}$ and a batch size of~$8$. 
Validation is performed every 200 iterations.
Attacks are performed using the network for which the validation loss is lowest.

\subsection{Evaluation}
We compare the drift reversal attack to related VoicePrivacy \emph{lazy-informed} and
\emph{semi-informed} attacks.
For the former, the attacker compares enrollment and trial utterances which are both in the $\postdom$ domain, but with an ASV model trained using data in the $\odom$ domain; other than by anonymizing the enrollment utterance, there is no compensation for operating upon anonymized data.
The
\emph{semi-informed} attacker makes greater effort and uses an ASV system that is trained using an independent set of similarly-anonymized data.
The latter is the default VoicePrivacy attack model.\footnote{It could be argued that drift reversal is also a \emph{semi-informed} attack, since it involves re-training a model on anonymized data (albeit unlabeled). However, for clarity, we use the term \emph{semi-informed} to refer to the attack method used in the VoicePrivacy Challenge 2022.}
The \emph{lazy-informed} attack is implemented using the original, pretrained ECAPA-TDNN for x-vector extraction.
The
\emph{semi-informed} attack is performed using an ECAPA-TDNN model which is fine-tuned using
AAM-softmax loss~\cite{aam_softmax}
and the same training settings as the drift reversal attack model.

Privacy evaluation results in terms of EER estimates are presented in \tabref{tab:attack} for each vocoder and each dataset.
EER results for unprotected data (no anonymization) are shown in column 3 and provide a reference against which EERs for protected data can be compared.
Results for the \emph{lazy-informed} attack are shown in column 4 and show substantial privacy gains (higher EERs).
This setting, however, gives a false sense of protection.
Results for the
\emph{semi-informed} attack shown in column~5 show considerably lower privacy gains; by retraining the ASV system using similarly anonymized data, the attacker can undo the anonymization to some degree. 
Results for the drift reversal attack also show universally lower privacy gains compared to the lazy-informed attack.  
These results add to the evidence that the role played by vocoder drift in anonymization is substantial and is also a potential weakness that can be exploited by an adversary.
The most powerful of the 3 attacks for each vocoder and dataset is highlighted in bold face and, for 5 of the 12 cases, the most powerful attack is drift reversal.

\section{Conclusions}
The work presented in this paper shows that, for the analyzed systems, the bulk of anonymization can be attributed not to the anonymization function of a conventional x-vector-based approach but, instead, to the vocoder function.
The cause is vocoder drift, namely the substantial difference between an input x-vector and the x-vector which can be extracted from the vocoder output.
This finding, while not necessarily surprising, calls into the question the research effort upon improving the anonymization function. One might wonder whether the design of different anonymization functions has any relevance at all, given that
the position of the final x-vector is dominated by the vocoder drift,
essentially nullifying the effort devoted to pseudo-speaker optimization.
This finding 
should not discourage further work in the design of x-vector anonymization functions, however.
Instead, it should encourage design toward
more grounded criteria.

Drift-reversal attacks rely on the fact that the x-vectors fed to the vocoder, though allegedly anonymized, still have a low level of protection. 
This is the result of an over-deterministic anonymization function; similar x-vector inputs will produce similar x-vector outputs, thus producing trial and enrollment speaker embeddings which are close in the output domain, and thus easy to match as the same speaker, even when anonymized. That is the case for the pool-based anonymization function.
Future work should investigate less deterministic anonymization functions to improve privacy directly in their output domain. Improvements to privacy in this domain will not only undoubtedly mitigate the risk of vocoder-drift-reversal attacks, but likely also that of \emph{semi-informed} attacks, which might inadvertently learn to exploit the same kind of vulnerability during training.

\balance
\bibliographystyle{IEEEtran}
\bibliography{mybib}

\end{document}